# Reply to "Comment on 'Scaling behavior of classical wave transport in mesoscopic media at the localization transition' "

S.K. Cheung and Z.Q. Zhang

Department of Physics, Hong Kong University of Science and Technology

Clear Water Bay, Kowloon, Hong Kong, China

16 Jun 2009

Abstract

We argue from both technical and physical points of view that the main result shown in the Comment by Cherrolet *et al.* [Phys. Rev. B **80**, 037101 (2009)] as well as the authors' interpretations of the result are not sufficient to draw the conclusion that the scaling law at the mobility edge takes the form $T \propto 1/L^2$. On the other hand, we believe that the result shows some evidence of $T \propto \ln L/L^2$ behavior found in S. K. Cheung and Z. Q. Zhang, Phys. Rev. B **72**, 235102 (2005). More calculations with even larger $L$'s are necessary to give a more definitive answer to this question.

In the preceding Comment (Ref. 1) on our Paper (Ref. 2), the authors fit their Eq. (1) to the average transmission coefficient, $T(L)$, of disordered slabs at the localization transition calculated by the self-consistent (SC) theory with a position-dependent diffusion constant, $D(z)$ (Ref. 3), in a range of slab thicknesses from $L=10^2 l$ to $8 \times 10^3 l$. Since deviations of the fit from the numerical results do not exceed 3% and Eq. (1) gives rise to $T \propto (l/L)^2$ behavior at large $L$s, they conclude that the $T \propto (l/L)^2 \ln(L/l)$ behavior we obtained in Ref. 2 was an artifact of replacing $D(z)$ with its harmonic mean. We would like to state here that Fig. 1 in the Comment as well as the authors' interpretations of this figure are not sufficient to draw a definitive conclusion about the scaling behavior of $T(L)$. Our reply is based on both technical and physical points of view.

On the technical side, we question the consistency and robustness in the determination of the parameter, $z_c = 4.2l$, in the assumed function, $D(z) = D(0)/(1+\tilde{z}/z_c)$ where $\tilde{z} = \min(z, L-z)$. In the Comment, the value of $z_c$ is determined from the fitting of Eq. (1) to the numerical transmission result, $T(L)$. Would $z_c$ be different if the fitting were done against the $D(z)$ obtained from the SC calculation? There are two reasons for us to raise this question. First, the value of $z_c=1.5l$ shown in the Table of Ref. 3 is very different from that found in the Comment. Second, in a standard form of $T$, the numerator in Eq. (8) of Ref. 3 takes the form $l + z_0$ (Ref. 4), where the term $l$ represents the penetration length, and, therefore, the numerator of Eq. (1) should be replaced by $4(z_c/l)(1+z_0/l)[D(0)/D_B]$. If we use this expression to fit $T(L)$, a different value of $z_c$ will be found. Thus, the claim of a good fit to within 3% might be ambiguous.

From the physical point of view, we argue that the critical behavior of $T(L)$ should be the large-$L$ behavior of $T$. If $T \propto (l/L)^2$ were the correct critical behavior as concluded in the Comment, the critical region of interest should be for $L > 1000l$, beyond which $T(L)(L/l)^2$ approaches a constant. However, such constant behavior is only seen in the region of $1000 < L/l < 3000$, which represents less than a decade of data points. $T(L)(L/l)^2$ turns into an increasing function of L for $3000 < L/l < 8000$, which indicates an overestimation of the localization effect by the assumed form of $D(z)$ in Eq. (1). In the Comment, the deviations from the constant behavior were explained as "*mostly due to the extremely slow convergence of our computational algorithm for thick slabs and would, most likely, disappear if more computer time were available.*" We believe, however, that there is physical reason behind the deviations and this physical reason is precisely the one that gives rise to the $T \propto (l/L)^2 \ln(L/l)$ behavior found in Ref. 2.

The reason for the overestimation of the localization effect is because the assumed form of $D(z) = D(0)/(1 + \tilde{z}/z_c)$ was obtained from a semi-infinite medium (Ref. 3). In any finite-sized sample, the decrease of $D(z)$ from a sample boundary should be slower than $D(0)/(1 + \tilde{z}/z_c)$ due to the presence of the other boundary that serves as a cutoff of the localization effect. This cutoff effect becomes more important in the middle region of the sample, where $D(z)$ is small. Since the transmission is dominated by the region of small $D(z)$, the absence of such a cutoff effect can lead to lower transmission as shown in Fig. 1 of the Comment. It is also the presence of the other boundary that gives rise to an upper cutoff length, $L$, for all diffusion modes along the z-axis, which, in turn, gives rise to the $T \propto \ln L / L^2$ scaling law (Ref. 2). Since the introduction of the position-dependent diffusion constant, $D(z)$, does not seem to remove this cutoff length, the $\ln L / L^2$ behavior is expected to be preserved in the SC calculation using $D(z)$ and to show up at larger $L$s. Thus, in our view, the deviations at large $L$s are evidence of the $T \propto \ln L / L^2$ scaling law. The above effect can be easily seen in one dimension. In a 1D semi-infinite medium, $D(z)$ is expected to decay exponentially (Ref. 3). Here, we adopt the 1D layered random media considered in Ref. 5. The dielectric constant in each layer is randomly distributed between 0.3 and 1.7. In this system, the transport mean free path, $l$, is almost identical to the localization length. We calculated $D(z)$ at a particular frequency for three sample thicknesses according to the method described in the Comment. The results are shown in Fig. 1. The long-dashed straight lines show the $D(z)$ obtained by using its exponentially decaying behavior in a semi-infinite medium with the reflection construction used in the Comment (Ref. 3). Such a construction clearly shows an overestimation of the localization effect in the middle region of each sample. While such overestimations are not expected to change the localization behavior in 1D, they may change the critical behavior of the 3D slabs considered here because of diverging correlation length.

Finally, we would like to point out that the difference between the result of Ref. 2 and numerical result shown in Fig. 1 of the Comment is mainly due to the use of different boundary conditions (BC) in obtaining the weak localization that appears in the term $\int_0^L dz [D_B / D(z)]$ in the denominator of Eq. (8) of Ref. 3. The use of harmonic-mean approximation in Ref. 2 is equivalent to the use of periodic BC. Since the assumed form of $D(z)$ in Eq. (1) is obtained previously from the fitting of some numerical $D(z)$ (Ref. 3), both the numerical result and Eq. (1) are obtained from the same mixed type BC. Thus, the good agreement between them shown in Fig. 1 of the Comment is not very surprising. This also explains the unphysical result of $T \propto (l/L)^2 \ln(L/\alpha l)$ with $\alpha \sim 10^{-24}$ obtained in the Comment as the fitting was done by using the results of two different BCs. While the mixed type BC is more physical than periodic BC, the good agreement shown in the range of $L$s considered might not necessarily be maintained when $L$ is further increased and surface effect diminishes. In the case of 2D Ising model, it has been shown that the critical scattering function can depend on the shape as well as BC of the system. However, the region that is specific to the surface effect shrinks with increasing sample size (Ref. 6).

In conclusion, the result presented in the Comment is not sufficient to draw the conclusion that the critical behavior at the mobility edge takes the form $T \propto 1/L^2$. On the other hand, we believe that the result presented here shows some evidence of $T \propto \ln L / L^2$ behavior found in Ref. 2. More calculations with even larger $L$s are necessary to give a more definitive answer to this question.

# References

[1] N. Cherroret, S.E. Skipetrov, and B.A. van Tiggelen, Preceding comment.

[2] S.K. Cheung and Z.Q. Zhang, Phys. Rev. B **72**, 235102 (2005).

[3] B.A. van Tiggelen *et al.*, , Phys. Rev. Lett. **84**, 4333 (2000).

[4] For example, J. G. Rivas *et al.*, Europhys. Lett. **48**, 22 (1999).

[5] Z.Q. Zhang *et al.*, Phys. Rev. B **79**, 144203 (2009).

[6] For example, P. Kleban *et al.*, Surface Science **166**, 159 (1986).

**Figure Caption**

FIG. 1. The function $D(z)$ obtained from the self-consistent calculation of a one-dimensional model considered in Ref. 5 at three sample thicknesses. The long-dashed straight lines show the $D(z)$ obtained by using its exponential decaying behavior in a semi-infinite medium with the reflection construction used in Refs. 1, 3.

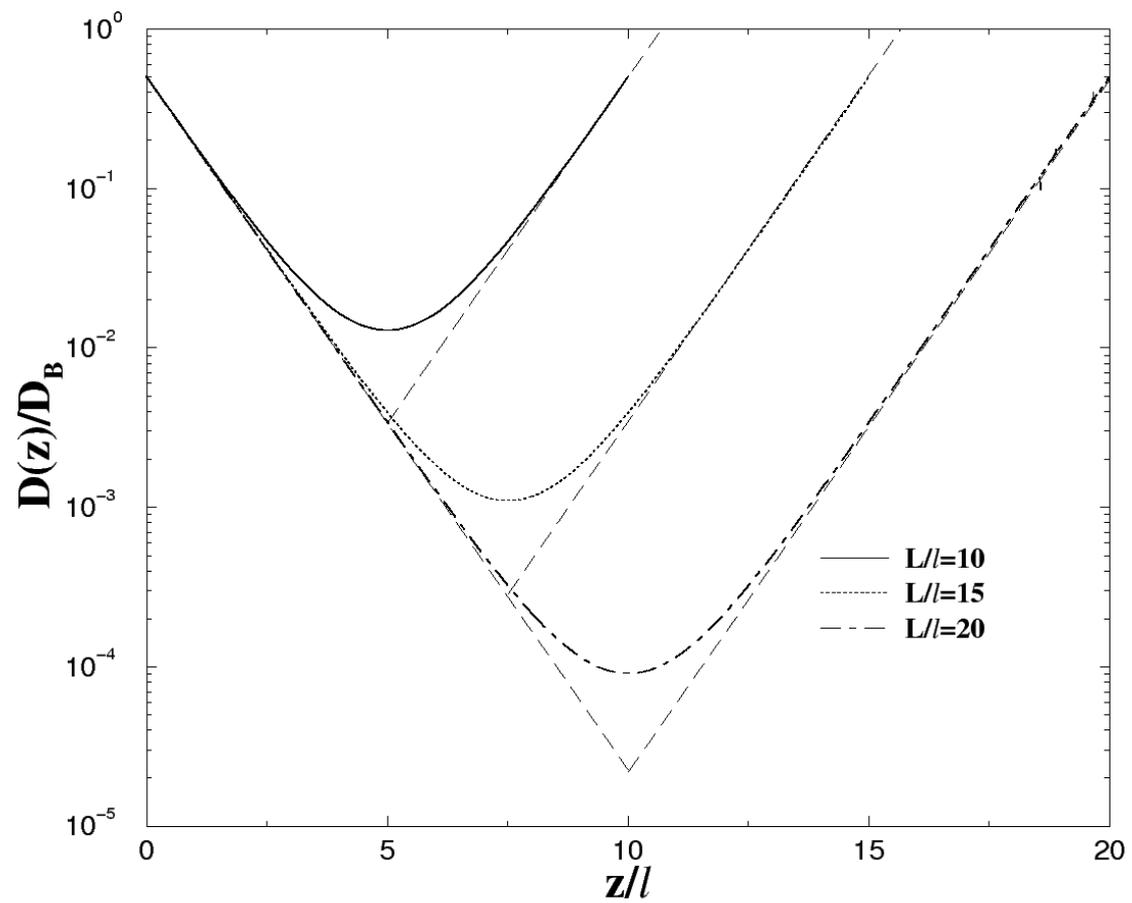

FIG. 1